\documentclass[11pt,a4paper]{article}

\usepackage[utf8]{inputenc}
\usepackage[T1]{fontenc}
\IfFileExists{lmodern.sty}{\usepackage{lmodern}}{}
\IfFileExists{microtype.sty}{\usepackage[expansion=false,protrusion=true]{microtype}}{}
\usepackage{amsmath,amssymb}
\usepackage{graphicx}
\usepackage{booktabs}
\usepackage{multirow}
\usepackage{enumitem}
\usepackage{float}
\usepackage{caption}
\usepackage{subcaption}
\usepackage[margin=2.4cm]{geometry}
\usepackage[colorlinks=true,linkcolor=blue,citecolor=blue,urlcolor=blue]{hyperref}
\graphicspath{{figures/}}
\newcommand{\ups}[1]{\ensuremath{\Upsilon(#1\mathrm{S})}}
\newcommand{\Ups}{\ensuremath{\Upsilon}}
\newcommand{\jpsi}{\ensuremath{J/\psi}}
\newcommand{\psit}{\ensuremath{\psi(2\mathrm{S})}}
\newcommand{\Ntrack}{\ensuremath{N_{\mathrm{track}}}}
\newcommand{\dNdeta}{\ensuremath{dN_{\mathrm{ch}}/d\eta}}
\newcommand{\pt}{\ensuremath{p_{\mathrm{T}}}}
\newcommand{\ST}{\ensuremath{S_{\mathrm{T}}}}
\newcommand{\Rtwentyone}{\ensuremath{R_{21}}}
\newcommand{\Rthirtyone}{\ensuremath{R_{31}}}

\title{A multi-differential constraint map for quarkonium suppression mechanisms in high-multiplicity \texorpdfstring{$pp$}{pp} and \texorpdfstring{$p$Pb}{pPb} collisions}
\author{Renato Campanini\\[2mm]
\small Dipartimento di Fisica e Astronomia, Universit\`a di Bologna,\\
\small and INFN, Sezione di Bologna, Italy\\
\small \texttt{renato.campanini@unibo.it}}
\date{}

\begin{document}
\maketitle

\begin{abstract}
The multiplicity-dependent suppression of $\Upsilon(nS)$ excited states
in high-multiplicity $pp$ and $p$Pb collisions is analysed using
publicly available CMS and LHCb data; preliminary CMS Physics Analysis
Summary results in $p$Pb and light-ion collisions are used only as
supporting cross-system evidence.
Six complementary differential constraints are considered: cone
isolation, azimuthal-sector equivalence, transverse sphericity,
transverse-momentum ordering,
forward-$E_T$ long-range correlation, and the pPb/Pbp forward-backward
asymmetry. Taken together, these constraints disfavour mechanisms
controlled solely by local track density or by total multiplicity, and
are consistent with an early, globally correlated, topology-sensitive
suppression pattern. The characteristic multiplicity scale at which suppression
sets in is independently consistent with the onset of a qualitative
change in soft-sector behaviour identified by Campanini and Ferri
\cite{CampaniniFerri2011} from inclusive charged-particle observables.
The result is a data-driven constraint map consistent with an early,
coloured pre-hadronic environment, possibly involving a deconfined stage.
\end{abstract}

\section{Introduction}

The suppression of quarkonium states is one of the classic probes of strongly interacting matter. In heavy-ion collisions, the sequential suppression of quarkonium bound states is commonly interpreted as the consequence of colour screening in a deconfined medium: the larger and more weakly bound states are suppressed first, while the ground state is more resistant to screening \cite{MatsuiSatz1986,Satz2006}. The observation of a multiplicity-dependent hierarchy among quarkonium states in proton--proton collisions therefore raises a sharply defined question: can such behaviour be generated entirely by hadronic final-state mechanisms in a small system, or does the differential structure of the data require additional early-time, non-local dynamics?

The present work addresses this question as a constraint analysis,
not as a new experimental measurement or as a new suppression model.
The aim is to determine which physical properties any successful
mechanism must possess in order to reproduce the complete set of
currently available differential measurements. In this sense the
analysis is deliberately model-discriminating rather than model-building.

The starting point is the CMS measurement of event-activity-dependent $\Upsilon(nS)$ production ratios in $pp$ collisions at $\sqrt{s}=7$ TeV \cite{CMS2020}. CMS measured not only the multiplicity dependence of the excited-to-ground-state ratios, but also their behaviour under several differential selections: track density inside a cone around the \Ups{} direction, azimuthal sectors relative to the \Ups{} direction, transverse sphericity, and transverse momentum. These observables separate local density, global activity, event topology and kinematic escape effects. The LHCb measurement of prompt \psit/\jpsi production in $pp$ collisions at $\sqrt{s}=13$ TeV \cite{LHCbPsi2024} provides an independent cross-check of the multiplicity and $p_T$ dependences. The LHCb measurement of $\Upsilon(nS)/\Upsilon(1S)$ ratios at $\sqrt{s}=13$ TeV \cite{LHCbUpsilon2025} provides an independent confirmation of the sequential suppression pattern at a different centre-of-mass energy, lending support to the robustness of the phenomenology. Two recent pPb results at $\sqrt{s_{_{\rm NN}}}=8.16$ TeV further extend the constraint map: the CMS Physics Analysis Summary HIN-25-005 (preliminary) on event-activity-dependent bottomonium production, including forward-$E_T$ and cone-isolation observables \cite{CMSPAS2025}, and the LHCb measurement of the prompt \psit/\jpsi ratio in pPb versus Pbp configurations \cite{LHCbpPb2025}. As further cross-system context, the recent preliminary CMS measurement
of sequential $\Upsilon(nS)$ suppression in symmetric light-ion
oxygen--oxygen and neon--neon collisions at
$\sqrt{s_{_{\rm NN}}}=5.36$~TeV \cite{CMSPAS2025OO} extends the
sequential hierarchy to an intermediate system size.  In the
single-ratio representation versus corrected track multiplicity, the
OO points follow the same smooth event-activity trend defined by the
published $pp$, $p$Pb and PbPb measurements, rather than forming a
disconnected system-specific pattern.  These preliminary pPb and light-ion results are used only as supporting
cross-system evidence. The core scissors constraint rests on the
published CMS $pp$ data, while the broader six-constraint map also uses
published LHCb and CMS pPb measurements where available.

The main result is a combined set of constraints. Local-density dissociation scenarios are constrained by the absence of a cone-density effect. Mechanisms depending only on total multiplicity are constrained by the strong sphericity dependence at fixed \Ntrack. The multiplicity dependence also weakens with increasing $\Upsilon$ $p_T$. Within the comover interaction model, both the local-density and partonic implementations of Ferreiro and Lansberg \cite{FerreiroLansberg2018} must be confronted with these constraints: the partonic CIM can better accommodate the cross-system density trend but is presently untested against the complete differential constraint set established here. The pPb data add two further constraints: the correlation of $\Upsilon$ suppression with forward $E_T$ across a large rapidity gap, and the contrasting behaviour of the $\psi(2S)/J/\psi$ ratio in pPb versus Pbp configurations. The resulting constraint set requires a mechanism that is not simply ``more particles produce more suppression''.

All numerical results used here are derived from publicly available experimental data and figures cited below. No new experimental measurement is claimed.

\section{Notation and observables}

The following notation is used throughout the paper. The variable \Ntrack{} denotes the CMS charged-track multiplicity in the acceptance $|\eta|<2.4$ with $\pt>0.4$ GeV/$c$. CMS provides the conversion between measured and corrected multiplicities and the corresponding charged-particle density. When a rough comparison with other small-system observables is needed, we use the CMS conversion $\Ntrack \simeq 2.83\,\dNdeta$ for the relevant selection, remembering that this is experiment- and cut-dependent rather than universal.

The quarkonium production ratios are
\begin{equation}
  R_{n1}=\frac{Y(\Upsilon(n\mathrm{S}))}{Y(\Upsilon(1\mathrm{S}))},\qquad n=2,3,
\end{equation}
where $Y$ denotes the yield in the relevant kinematic selection.
No double-excited-state ratio is used in the present analysis; the
sequential hierarchy is assessed directly from the separate behaviour
of $R_{21}$ and $R_{31}$.

The cone-isolation variable $N^{\Delta R}_{\rm track}$ counts tracks inside a cone $\Delta R<0.5$ around the \Ups{} direction, with $\Delta R=\sqrt{(\Delta\eta)^2+(\Delta\phi)^2}$. The azimuthal-sector multiplicities are defined relative to the \Ups{} direction: forward $|\Delta\phi|<\pi/3$, transverse $\pi/3<|\Delta\phi|<2\pi/3$, and backward $|\Delta\phi|>2\pi/3$. The transverse sphericity $\ST=2\lambda_2/(\lambda_1+\lambda_2)$ is obtained from the eigenvalues of the transverse-momentum tensor; $\ST\to0$ is jet-like and $\ST\to1$ is isotropic.

\section{Multiplicity dependence and transverse-momentum ordering}
\label{sec:ptordering}

CMS observes a monotonic decrease of both \Rtwentyone{} and \Rthirtyone{} with \Ntrack{} in $pp$ collisions at $\sqrt{s}=7$ TeV, in both the inclusive sample ($p_T^{\mu\mu}>0$~GeV, 0.3~fb$^{-1}$) and the high-$\pt$ sample ($p_T^{\mu\mu}>7$~GeV/$c$, 4.8~fb$^{-1}$). The effect is stronger for \ups{3} than for \ups{2}, as expected for sequential suppression ordered by binding energy.

\begin{figure}[H]
\centering
\includegraphics[width=0.95\textwidth]{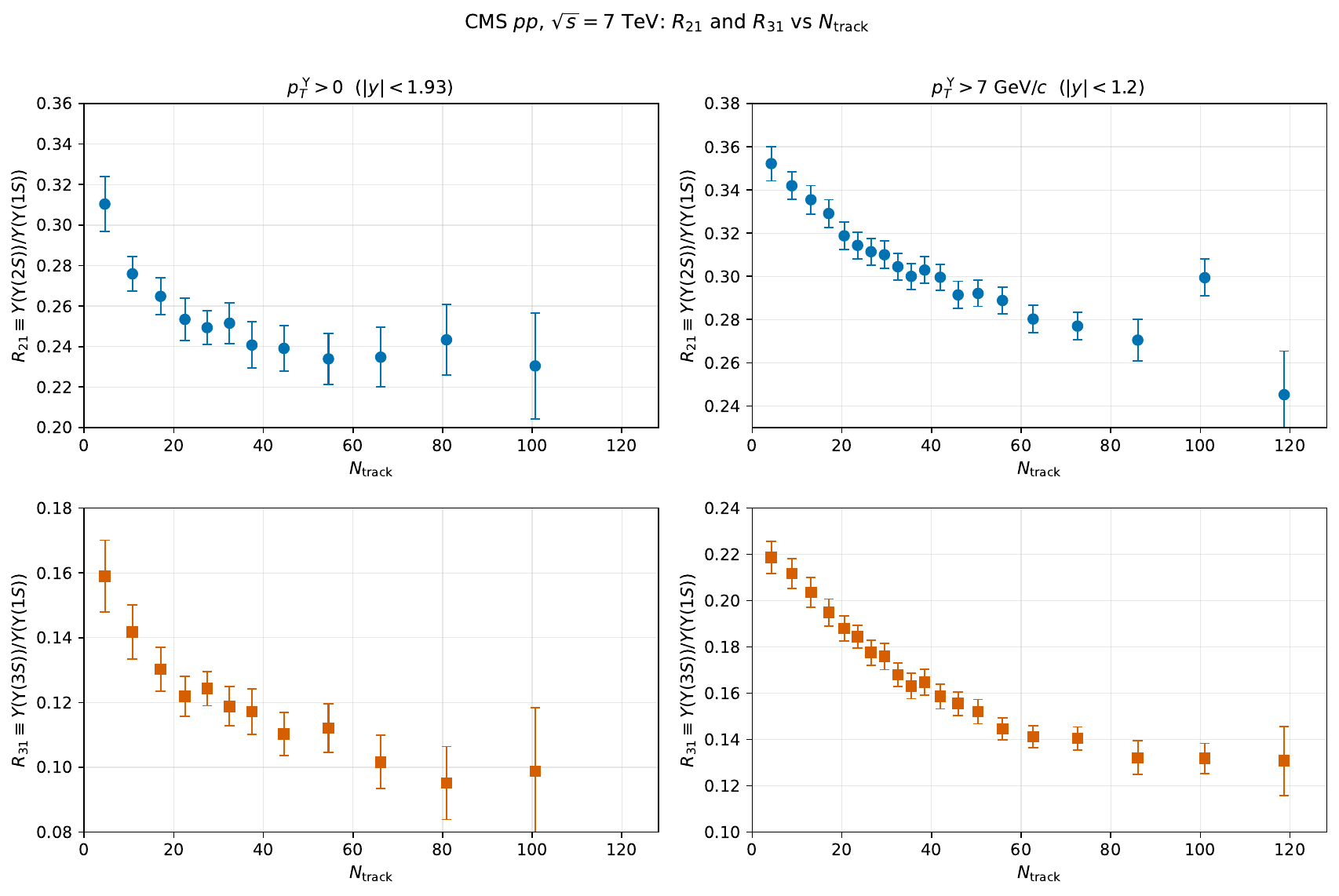}
\caption{CMS $pp$ data at $\sqrt{s}=7$ TeV: \Rtwentyone{} and \Rthirtyone{} as functions of \Ntrack{} in inclusive and high-$\pt$ selections. The figure is reproduced from the public CMS/HEPData-based plot used for the present analysis. Both ratios decrease with multiplicity, with a stronger relative decrease for the more weakly bound \ups{3} state.}
\label{fig:cms_ratios}
\end{figure}

The transverse-momentum dependence provides an additional constraint. At fixed \Ntrack, higher-\pt{} quarkonia are less suppressed. In the CMS \Ups{} data the multiplicity slope weakens as the \Ups{} transverse momentum increases, and the highest-$\pt$ slice is compatible with a much weaker multiplicity dependence within uncertainties. This is the qualitative behaviour expected from a finite-size medium: a fast quarkonium crosses the active region on a shorter timescale,
\begin{equation}
  \tau_{\rm cross}\sim \frac{R}{\beta\gamma},
\end{equation}
so its survival probability increases with momentum~\cite{GunionVogt1996}.

\begin{figure}[H]
\centering
\includegraphics[width=0.92\textwidth]{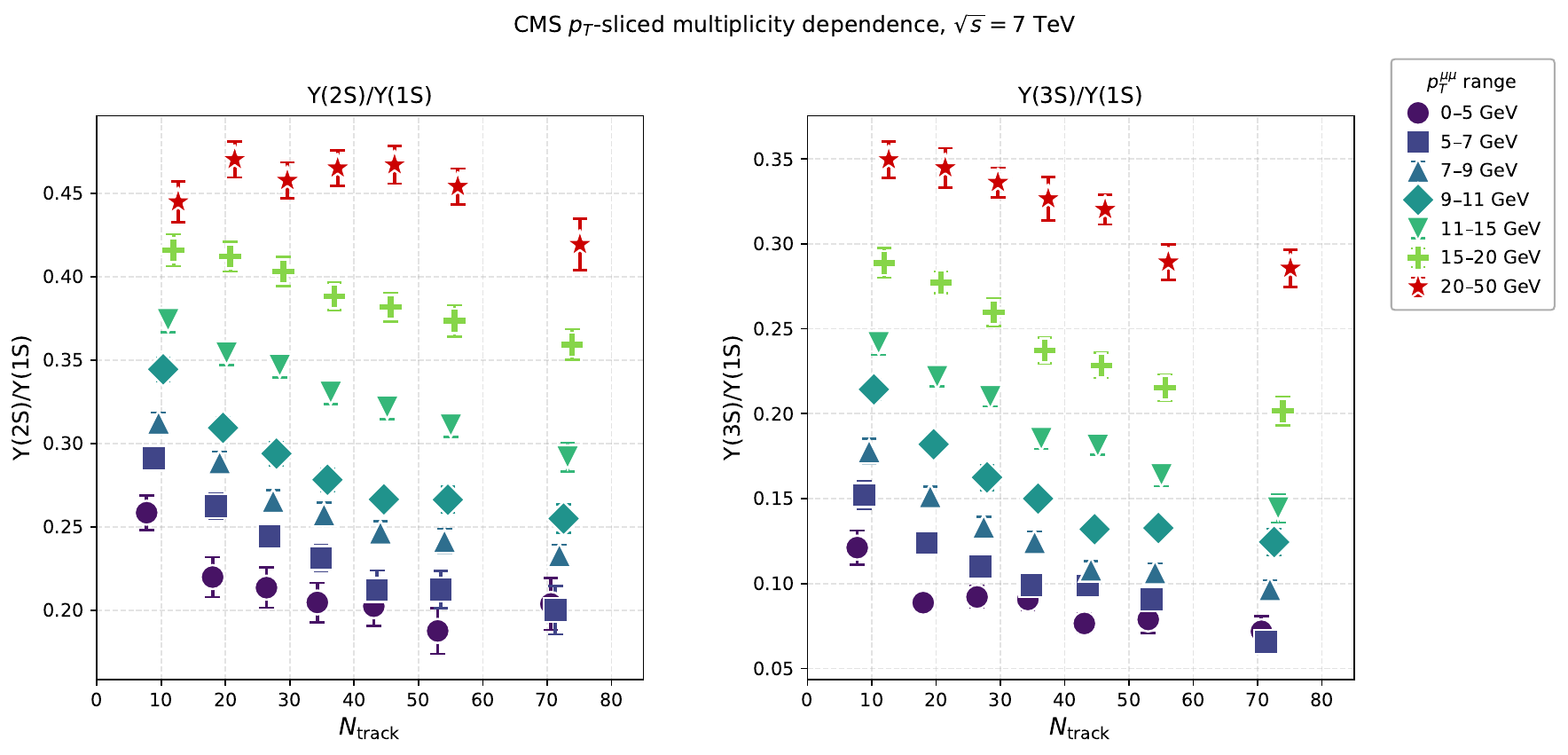}
\caption{CMS $p_T$-sliced multiplicity dependence of \Rtwentyone{} and \Rthirtyone{} in $pp$ collisions at $\sqrt{s}=7$ TeV. The suppression is strongest in low-$p_T$ slices and weakens at high $p_T$, consistent with a finite-size escape effect.}
\label{fig:pt_slices}
\end{figure}

The mean transverse momentum of the surviving \Ups{} population carries the same information in a complementary form. At high multiplicity, the surviving-state mean-$p_T$ hierarchy
\begin{equation}
\langle p_T\rangle_{\Upsilon(3S)} >
\langle p_T\rangle_{\Upsilon(2S)} >
\langle p_T\rangle_{\Upsilon(1S)}
\end{equation}
is observed in the CMS data \cite{CMS2020}.  This is qualitatively
consistent with a survival-bias picture in which the fragile excited
states lose a larger fraction of their low-$p_T$ component, shifting
the surviving population to higher mean transverse momentum.

\begin{figure}[H]
\centering
\includegraphics[width=0.92\textwidth]{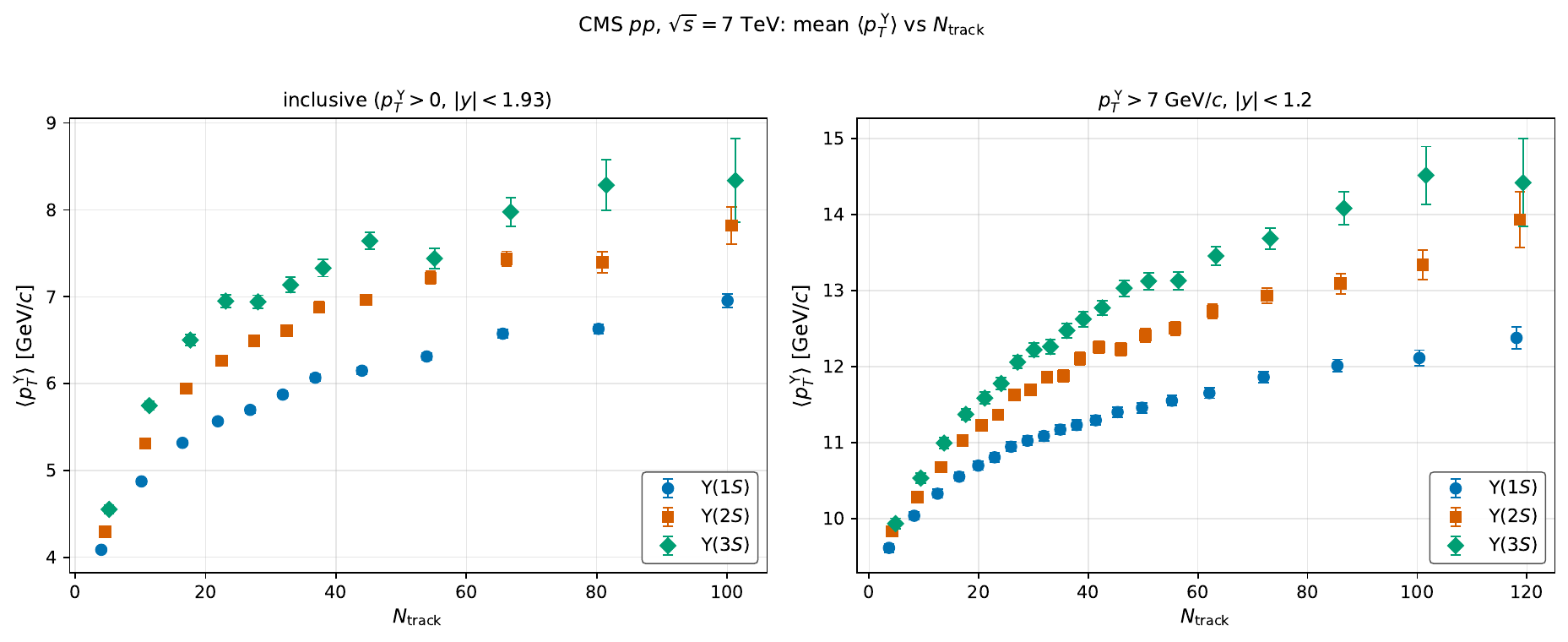}
\caption{Mean transverse momentum of the three \Ups{} states as a function of \Ntrack{} in CMS $pp$ data. The high-multiplicity hierarchy of the surviving population is consistent with preferential depletion of low-$p_T$ excited states.}
\label{fig:meanpt}
\end{figure}

The multiplicity trend and the transverse-momentum ordering are not by themselves sufficient to discriminate among suppression scenarios. The discriminating information comes from the geometry, topology,
transverse-momentum dependence, and long-range event-activity structure
of the event.

\section{Combined constraint framework}
\label{sec:combined}

The analysis is organised around six complementary constraints:
\begin{enumerate}
  \item \textbf{Cone-multiplicity independence:} the suppression ratios
        show no significant dependence on the charged-particle
        multiplicity inside a cone $\Delta R < 0.5$ around the
        $\Upsilon$ direction, indicating that local track density near
        the quarkonium is not the controlling variable.
  \item \textbf{Azimuthal independence:} the suppression trends are
        statistically compatible when measured using forward, transverse,
        or backward sector multiplicity, showing no preferred direction
        relative to the quarkonium.
  \item \textbf{Transverse-momentum dependence:} the multiplicity
        dependence of the suppression weakens progressively with
        increasing $\Upsilon$ $p_T$.
  \item \textbf{Sphericity and topology dependence:} at fixed
        \Ntrack{}, jet-like and isotropic events show different
        suppression patterns, demonstrating that event topology carries
        independent physical information beyond total multiplicity.
  \item \textbf{Forward-$E_T$ long-range correlation:} CMS already
        observed in published $p$Pb data at $\sqrt{s_{_{\rm NN}}}=5.02$~TeV
        that $\Upsilon(nS)/\Upsilon(1S)$ ratios decrease with the forward
        transverse energy $E_T(|\eta|>4)$, measured several units of
        pseudorapidity away from the quarkonium \cite{CMS2014pPb}. The
        preliminary 8.16~TeV result \cite{CMSPAS2025} confirms and extends
        this long-range activity correlation with higher statistics.
  \item \textbf{pPb/Pbp forward-backward asymmetry:} the prompt
        $\psi(2S)/J/\psi$ ratio decreases with multiplicity in the
        proton-going direction, while in the lead-going direction it is
        suppressed but approximately flat; any viable mechanism must
        reproduce both behaviours simultaneously.
\end{enumerate}
Any successful description of these data must address all of these
constraints simultaneously.
This combined requirement is more restrictive than any individual
observable.

\section{Cone isolation: constraint on local-density dissociation}
\label{sec:cone}

CMS classifies events according to the track density inside a cone $\Delta R<0.5$ around the \Ups{} flight direction, and presents the \Rtwentyone{} and \Rthirtyone{} ratios as functions of \Ntrack{} separately for four cone categories, $N^{\Delta R}_{\rm track}=0,1,2$, and $>2$. The two limiting classes are an empty cone, $N^{\Delta R}_{\rm track}=0$, and a dense cone, $N^{\Delta R}_{\rm track}>2$.

For a local hadronic dissociation mechanism, the expected sign is unambiguous: more nearby hadrons should increase the dissociation probability. The data do not show such a separation: the multiplicity-dependent trends of the empty-cone and dense-cone
categories are statistically compatible within the present uncertainties
across the measured \Ntrack{} range.  Any difference between the
empty-cone and dense-cone trends is smaller than the experimental
uncertainties.  No significant excess suppression associated with
enhanced local charged-particle activity is observed.  The present result
should therefore be interpreted as a constraint on large local-density
effects rather than as a proof that local interactions are entirely absent.

\begin{figure}[H]
\centering
\includegraphics[width=0.92\textwidth]{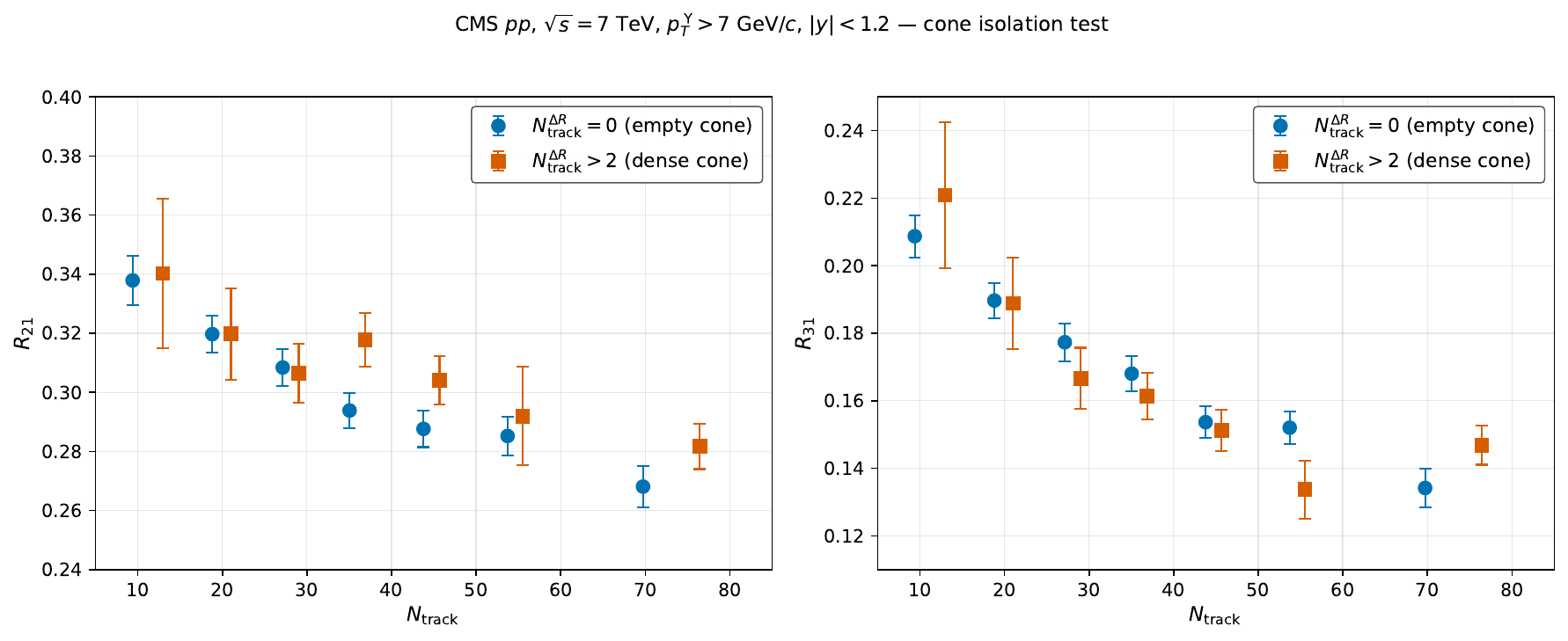}
\caption{Cone-isolation test in CMS $pp$ data at $\sqrt{s}=7$ TeV. Empty-cone and dense-cone selections give statistically compatible \Rtwentyone{} and \Rthirtyone{} values. This places direct pressure on mechanisms in which the suppression probability scales with the local density of hadrons around the quarkonium.}
\label{fig:cone}
\end{figure}

The conclusion is not that every conceivable comover-type construction is excluded. Rather, the local-density interpretation of the comover picture is challenged by this result: the variable controlling the suppression cannot simply be the number of charged particles close to the reconstructed \Ups{} direction. No published comover implementation has yet been shown to reproduce
this cone-isolation constraint together with the other differential
constraints considered here.

\section{Azimuthal sectors: constraint on directional mechanisms}
\label{sec:azimuth}

The cone result could still leave room for a more diffuse directional mechanism. CMS addresses this by measuring the suppression ratios as functions of the multiplicity in three azimuthal sectors relative to the \Ups{} direction: forward, transverse and backward.

\begin{figure}[H]
\centering
\includegraphics[width=0.92\textwidth]{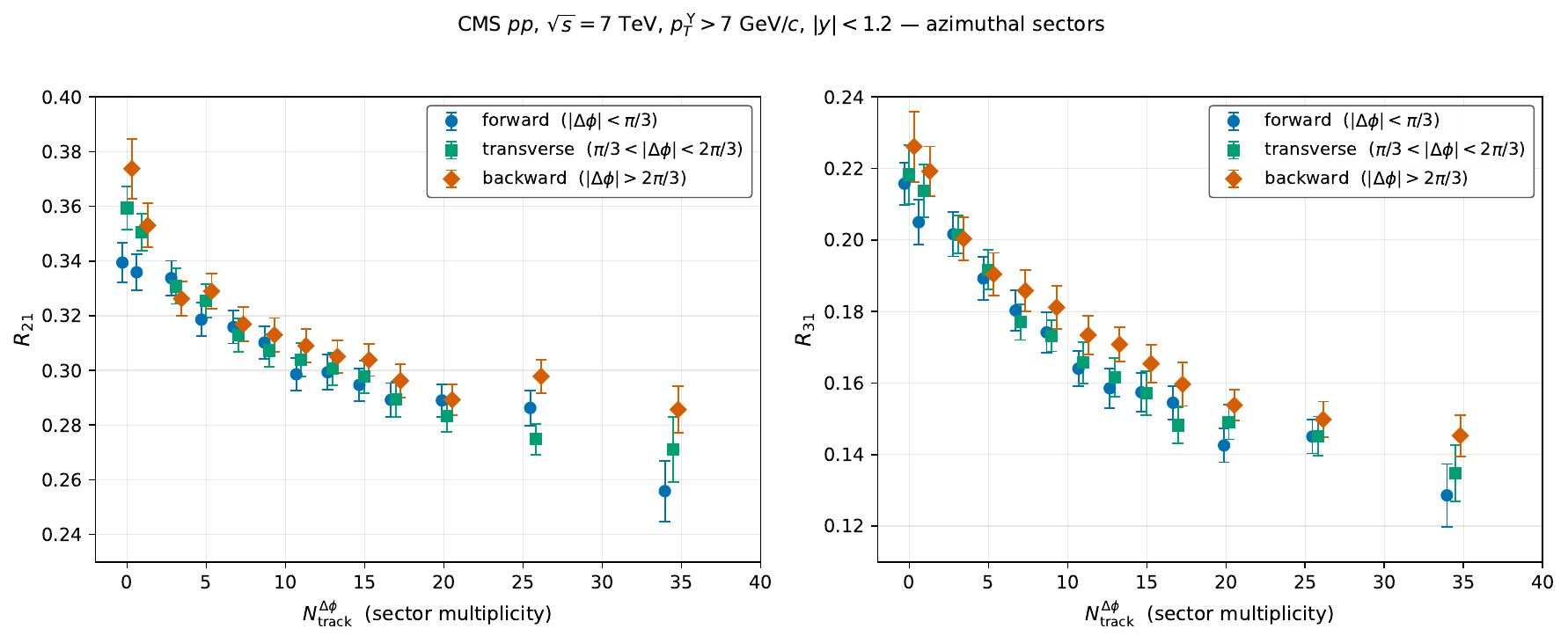}
\caption{CMS azimuthal-sector test. The suppression trends obtained using forward, transverse and backward sector multiplicities are mutually compatible within uncertainties. The transverse sector, located around $90^\circ$ relative to the \Ups{} direction, is especially constraining because it is neither co-moving with nor recoiling against the quarkonium.}
\label{fig:azimuth}
\end{figure}

The transverse sector is particularly informative. Tracks at roughly $90^\circ$ relative to the \Ups{} direction have no natural interpretation as local comovers or recoil partners, yet their multiplicity correlates with the suppression in the same way as forward and backward tracks. Thus the sector multiplicity acts as a proxy for global event activity
rather than as a directional local density.  No statistically significant
hierarchy among forward, transverse and backward sector trends is observed;
all three are mutually compatible within uncertainties.  Direction-based
and near-side explanations are therefore disfavoured in their simplest forms.

\section{Transverse sphericity: constraint on multiplicity-only mechanisms}
\label{sec:sphericity}

If the cone and azimuthal data indicate that the relevant variable is not local density, a natural rescue is to assume that the suppression depends only on the global event activity \Ntrack. The transverse-sphericity measurement tests exactly this possibility. CMS measures the $\Upsilon(nS)/\Upsilon(1S)$ ratios as functions of \Ntrack{} separately in four transverse-sphericity intervals spanning the full range $0<\ST<1$. We focus here on the two extreme classes, jet-like events ($\ST<0.55$) and isotropic events ($\ST>0.85$), at the same \Ntrack, since these show the most significant separation; a quantitative comparison with the $p_T$-sliced data is given in Appendix~\ref{app:slopes}.

\begin{figure}[H]
\centering
\includegraphics[width=0.92\textwidth]{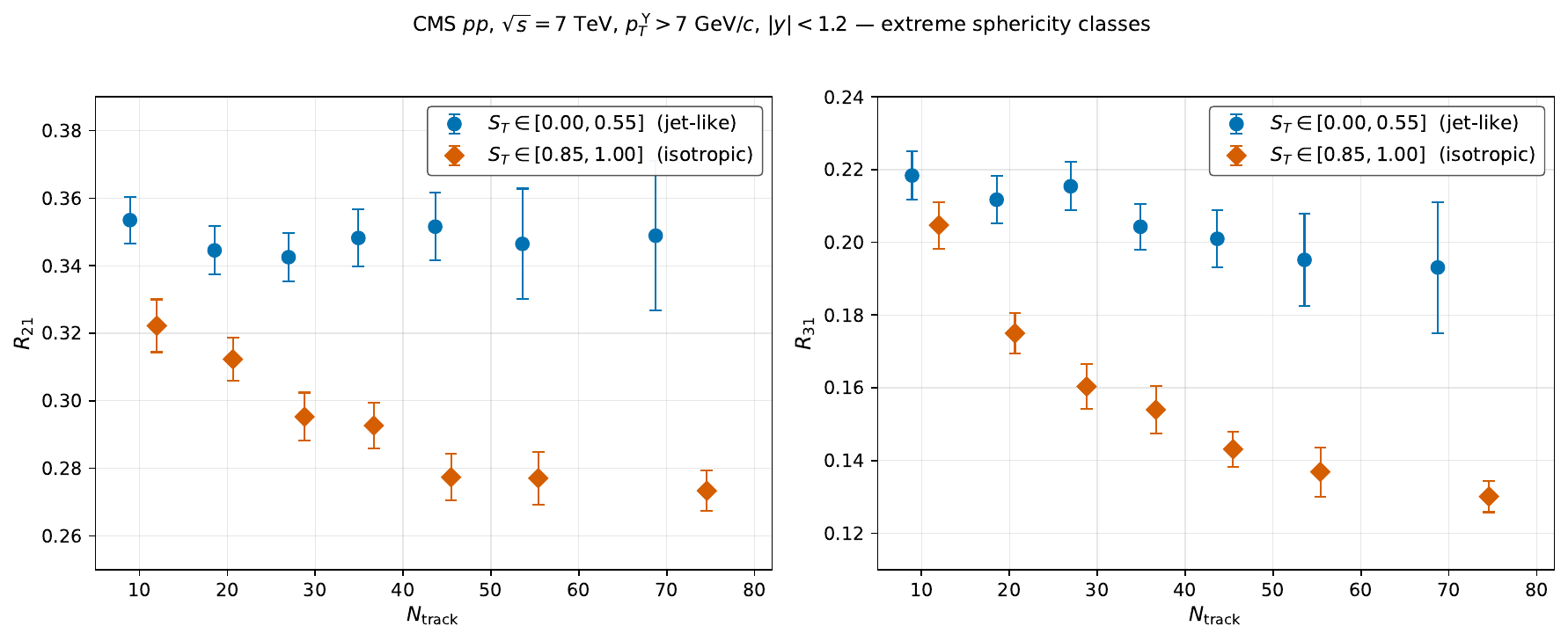}
\caption{CMS sphericity test. Jet-like and isotropic events with comparable \Ntrack{} show different suppression patterns. This demonstrates that the suppression is not a function of \Ntrack{} alone and that event topology contains independent physical information.}
\label{fig:sphericity}
\end{figure}

The observed topology dependence is the key discriminator.  As reported
by CMS \cite{CMS2020}, in jet-like events ($S_T < 0.55$) the ratios are
approximately independent of \Ntrack{}, while in isotropic events
($S_T > 0.85$) both \Rtwentyone{} and \Rthirtyone{} decrease with
multiplicity.  CMS explicitly notes that the multiplicity dependence
of the ratios is present for $S_T > 0.55$ and largely absent in the
jet-like class, and concludes that the decrease is an underlying-event
effect.  The two topology classes therefore carry independent physical
information beyond total multiplicity: a mechanism depending only on
\Ntrack{} would produce the same multiplicity dependence in both
subsamples, which the data do not support.

A second, simpler observable in the same CMS paper makes the same
point even more starkly, independently of the ratio formalism used
above. Inclusively, the \Ups{} states are accompanied by different
mean track multiplicities: $\langle \Ntrack\rangle = 33.9\pm0.1$ for
\ups{1}, $33.0\pm0.1$ for \ups{2}, and $32.0\pm0.1$ for \ups{3}
\cite{CMS2020}. Within the jet-like class alone ($0<S_T<0.55$),
however, CMS finds this difference to vanish entirely: the mean
associated multiplicity is identical for all three states,
$\langle\Ntrack\rangle=22.4\pm0.1$ \cite{CMS2020}. CMS concludes from
this that the state-dependent difference in associated multiplicity is
not directly linked to the mass difference between the three states,
but is instead an effect tied to the isotropic (underlying-event)
topology. This single number --- an exact equality across three
otherwise distinct particle states once jet activity is removed ---
is one of the cleanest demonstrations in the dataset that the
multiplicity-state correlation is a topology effect rather than an
intrinsic property of each \Ups{} state.

The topology dependence is not limited to a comparison between the two
extreme event classes.  The suppression strength of $R_{31}$ evolves
approximately progressively across the four measured $S_T$ intervals:
the total variation $|\Delta R_{31}|$ across the $\Ntrack$ range grows
from 0.025 in the most jet-like class ($0 < S_T < 0.55$) to 0.055,
0.058, and 0.075 in the progressively more isotropic classes.
This indicates that transverse sphericity acts as a continuous
event-shape variable rather than as a binary selection, and makes
explanations based on an accidental difference between the two extreme
subsamples less natural; a quantitative comparison of the jet-like
slopes against the published $p_T$-sliced measurements is given in
Appendix~\ref{app:slopes}.

Together with the cone result, this produces a ``scissors'' constraint.
Formally: cone isolation shows that the multiplicity-dependent trend of
the suppression ratios is statistically indistinguishable across
cone categories,
\begin{equation}
R_{n1}(\Ntrack\,|\,N^{\Delta R}_{\rm track}) \approx
R_{n1}(\Ntrack),
\end{equation}
while sphericity shows that
\begin{equation}
R_{n1}(\Ntrack\,|\,S_T) \neq R_{n1}(\Ntrack).
\end{equation}
These two observations are orthogonal: cone isolation probes whether
the multiplicity-dependent suppression trend is sensitive to the local
track density around the quarkonium; sphericity probes whether, for a
given multiplicity dependence, the trend itself differs with event
topology.
Their simultaneous occurrence strongly constrains any mechanism whose suppression
probability is determined solely by a single event variable.
Any successful model must therefore be global but not merely multiplicity-driven;
it must be sensitive to the topology or energy-density structure of the event.
This pair of observations constitutes the central experimental result
of the present work: it simultaneously disfavours suppression mechanisms
controlled exclusively by local particle density and those controlled
exclusively by total multiplicity.

\subsection*{The sphericity dependence is unlikely to be explained by a $p_T$ composition bias}

A natural objection is whether the observed slope difference between
jet-like and isotropic events could be explained by the different
$p_T$ compositions of the two samples: jet-like events tend to have
harder $\Upsilon$ transverse-momentum spectra, and the $p_T$ ordering
of Section~\ref{sec:ptordering} shows that high-$p_T$ quarkonia are less
suppressed.

The presently available CMS measurements do not permit a rigorous
separation of topology and transverse-momentum effects, because the
$p_T$-sliced and sphericity-sliced analyses are performed on different
projections of the same dataset: the $p_T$-sliced data are integrated
over all $S_T$, and the sphericity-sliced data are integrated over all
$p_T$.
A triple-differential measurement $R_{n1}(N_{\rm track},\,S_T,\,p_T)$
would be required for a definitive decomposition; this is identified as
a priority follow-up (Section~\ref{sec:predictions}).

Nevertheless, the existing projections are not naturally consistent
with a purely kinematic interpretation.  CMS explicitly reports that
in all measured $p_T^{\mu\mu}$ ranges the ratios decrease with
increasing multiplicity \cite{CMS2020}: the dependence weakens at
high $p_T$ but does not vanish, and no published $p_T$ slice reproduces
the near-flat behaviour of the jet-like class.  A quantitative
comparison between the $p_T$-sliced and sphericity-sliced slopes,
using weighted fits to the CMS HEPData central values, is given in
Appendix~\ref{app:slopes}; it shows that no single $p_T$ interval
simultaneously reproduces both the normalisation and the multiplicity
dependence of the jet-like class.

In summary: we explicitly refrain from claiming that the topology
dependence is independent of the transverse-momentum composition ---
the present data do not permit such a separation without the
triple-differential measurement $R_{n1}(N_{\rm track},S_T,p_T)$
proposed in Section~\ref{sec:predictions}.
Nevertheless, the existing $p_T$-sliced and $S_T$-sliced projections
do not naturally support a purely kinematic origin for the observed
sphericity dependence: the $p_T$ composition may contribute to the
splitting, but cannot by itself account for the near-vanishing slope
of the jet-like class, which lies outside the range of every published
$p_T$ slice.
The topology dependence is therefore treated here as an independent
differential constraint, with the explicit caveat that a definitive
decomposition awaits the triple-differential measurement.
A quantitative comparison of these slopes with the published $p_T$-sliced
measurements is given in Appendix~\ref{app:slopes}.

\section{LHCb charmonium: multiplicity and \texorpdfstring{$p_T$}{pT} dependence}
\label{sec:temporal}

The LHCb measurement of the prompt $\psi(2S)/J/\psi$ ratio in
$pp$ collisions at $\sqrt{s}=13$~TeV \cite{LHCbPsi2024} provides an
independent cross-check of the multiplicity and transverse-momentum
dependences observed in bottomonium.

The prompt $\psi(2S)/J/\psi$ ratio decreases with event activity, and
LHCb rejects the hypothesis of a constant ratio with high significance
\cite{LHCbPsi2024}.  Its multiplicity dependence also weakens
progressively with increasing $p_T$, approaching compatibility with a
flat behaviour in the highest-$p_T$ interval.  This ordering is
consistent with the finite-size interpretation already suggested by
the CMS bottomonium $p_T$ slices: a high-$p_T$ quarkonium traverses any
finite active region in a shorter proper time and is therefore less
affected by the suppressing environment.

The non-prompt $\psi(2S)/J/\psi$ ratio is flat within uncertainties
across multiplicity and $p_T$ intervals.  This observation is useful as
a consistency check that the multiplicity dependence is associated
with the primary-collision environment rather than with delayed
$b$-hadron decays.  However, it provides only a coarse temporal
separation and is not used here as an independent timing constraint on
the microscopic suppression mechanism.  The more relevant early-time
argument is instead the formation-time consideration discussed in
Section~\ref{sec:prehadronic}.

\begin{figure}[H]
\centering
\includegraphics[width=0.92\textwidth]{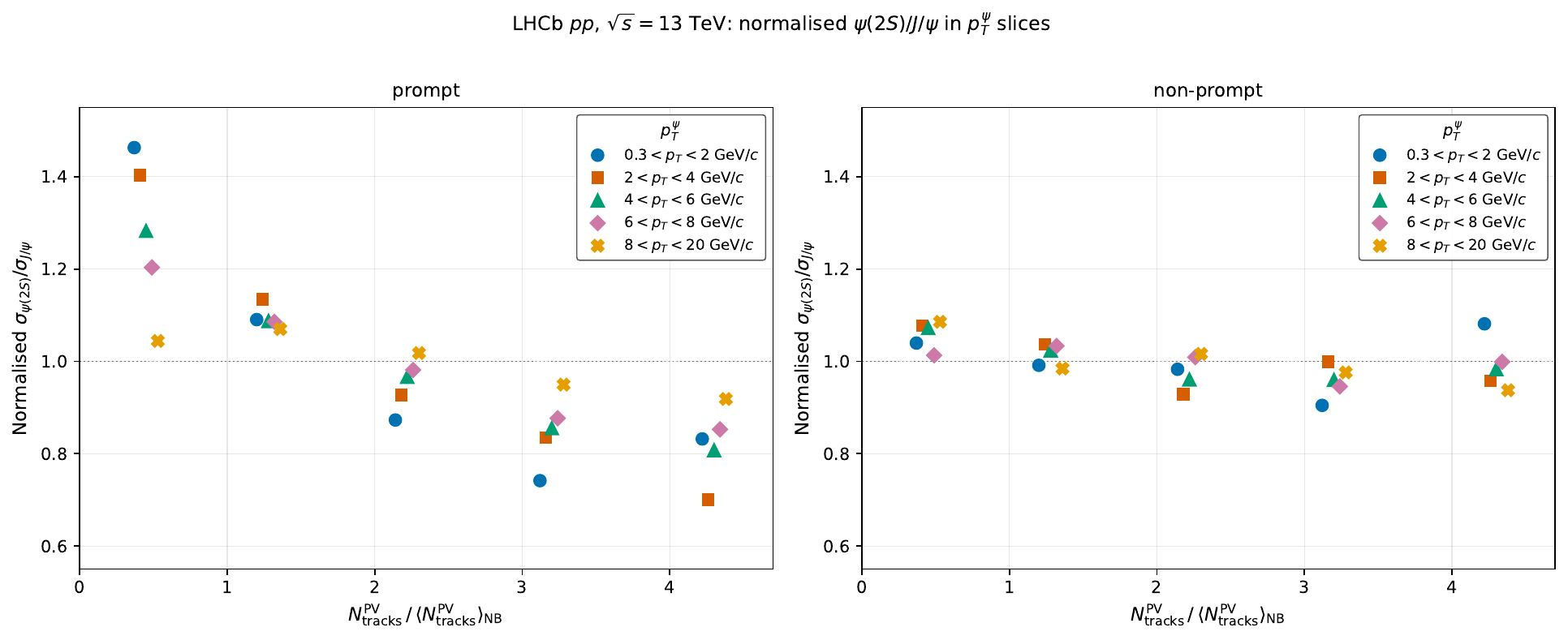}
\caption{%
Normalised prompt (left panels) and non-prompt (right panels)
$\psi(2\mathrm{S})/J/\psi$ production ratio as a function of
charged-particle multiplicity $\Ntrack/\langle\Ntrack\rangle_{\rm MB}$
in $pp$ collisions at $\sqrt{s}=13$~TeV, shown in five $p_T$ intervals.
Data extracted from LHCb HEPData \cite{LHCbPsi2024}.
The non-prompt ratio is consistent with a constant (dashed line) in
every $p_T$ bin.
The prompt ratio shows a suppression that weakens with increasing $p_T$,
consistent with a finite-size escape effect.
Error bars represent the published statistical and systematic
uncertainties added in quadrature.}
\label{fig:lhcb_ptslices}
\end{figure}

\section{Constraints on comover models and partonic extensions}
\label{sec:models}

The combined constraints can be used to assess the main suppression
frameworks systematically.

\paragraph{Comover interaction model (CIM).}
In the CIM \cite{ArmestoCapella1998,GavinVogt1997} the suppression
probability scales with the local comover density at the rapidity and
transverse position of the quarkonium pair.  The cone and azimuthal
data (Constraints~1--2) show no excess suppression in dense-cone or
in any particular azimuthal sector, directly constraining this
local-density picture.  Replacing local with global event-averaged
density resolves Constraints~1--2 but not Constraint~4 (sphericity):
isotropic and jet-like events at fixed \Ntrack{} must then give the
same suppression, which the data do not support.

The partonic CIM of Ferreiro and Lansberg \cite{FerreiroLansberg2018}
operates at $\tau_0\sim0.1$~fm, well before hadronisation, and
reproduces minimum-bias pPb and PbPb suppression.  To our knowledge, it has not been
tested against any of the six differential constraints of the present
paper.  Extending it to cone-selected, $p_T$-sliced, sphericity-resolved, and
forward-$E_T$ samples is the most direct test available.  The CIM
remains a viable candidate; its compatibility with the full constraint
set is an open question.

\paragraph{Medium-based approaches: OQS/pNRQCD and QGP-droplet models.}
Frameworks based on open-quantum-system evolution derived from
pNRQCD~\cite{Brambilla2011,Strickland2024} have been applied successfully
to quarkonium suppression in large systems and provide a natural
theoretical language for finite-size medium effects; they have not,
to our knowledge, been tested against any of the six differential
constraints established here.  The QGP droplet model of Bai and
Chen~\cite{ChenQGP} is designed for small systems and reproduces
charmonium suppression in $pp$; the preliminary CMS HIN-25-005
\cite{CMSPAS2025} finds qualitative agreement with a QGP-based
framework in pPb, in contrast to the comover model.  The six
differential constraints are open tests for both classes.

\paragraph{Hydrodynamic dissociation: SHINCHON.}
The SHINCHON model \cite{SHINCHON2023} combines event-by-event
hydrodynamics with bottomonium dissociation including feed-down, and
has been compared by CMS to OO and PbPb double ratios
\cite{CMSPAS2025OO}.  To our knowledge, it has not been confronted with any of the six
differential constraints.

\paragraph{Universal requirement.}
All six differential constraints must be simultaneously reproduced by
any viable suppression mechanism.  To our knowledge, no published
framework has yet been tested against this full set.  The present work establishes this
experimental benchmark; future calculations will determine which
mechanisms satisfy it.

\section{Cross-experimental robustness}
\label{sec:lhcb13}

\subsection{LHCb \texorpdfstring{$\Upsilon(nS)$}{Upsilon(nS)} at 13 TeV}

An independent confirmation of the sequential suppression pattern is
provided by the LHCb measurement of $\Upsilon(nS)/\Upsilon(1S)$ ratios
as a function of event activity in $pp$ collisions at $\sqrt{s}=13$~TeV
\cite{LHCbUpsilon2025}.  This measurement uses a different detector,
a different centre-of-mass energy, and a different rapidity coverage
($2<y<4.5$) than the CMS data.  The same qualitative features are
observed: a decreasing trend of $\Upsilon(2S)/\Upsilon(1S)$ and
$\Upsilon(3S)/\Upsilon(1S)$ with increasing event activity, with the
more weakly bound state showing a larger suppression.

The LHCb data also show a $p_T$ dependence consistent with
Constraint~3: at high $\Upsilon$ transverse momentum the multiplicity
dependence of the ratios declines, in agreement with the trend
observed in CMS~7~TeV data \cite{CMS2020} and with the ATLAS
associated-multiplicity measurement at 13~TeV \cite{ATLASCONF2022023}.
Importantly, even in the hardest available bin
($10 < p_T < 30$~GeV/$c$) the suppression remains clearly present:
both $\Upsilon(2S)/\Upsilon(1S)$ and $\Upsilon(3S)/\Upsilon(1S)$
decrease substantially from low to high multiplicity.
This stands in contrast to the behaviour observed in jet-like events
in CMS, where selecting $S_T < 0.55$ causes the multiplicity dependence to
largely disappear.  High $p_T$ therefore weakens the
multiplicity dependence but does not remove it, while a jet-like
topology selection largely removes it.

The fact that the same hierarchy and multiplicity dependence appear in
two independent experiments at two different energies and rapidities
strengthens the case that the effect is a genuine physical phenomenon
rather than an experimental artefact.

\subsection{ATLAS associated multiplicity and feed-down cross-check}
\label{sec:atlasue}

A useful cross-experimental check is provided by the ATLAS measurement
of the charged-particle multiplicity associated with $\Upsilon(nS)$
production at $\sqrt{s}=13$~TeV \cite{ATLASCONF2022023}. This
ATLAS-CONF-2022-023 is a full Run~2 result (139~fb$^{-1}$), an
official collaboration-approved document that has already been cited by
subsequent published work including the LHCb measurement of
Section~\ref{sec:lhcb13} \cite{LHCbUpsilon2025}. ATLAS measures the
same class of observable used by CMS in Section~\ref{sec:sphericity}:
the event activity accompanying each bottomonium state. The difference
is that ATLAS studies this observable as a function of the $\Upsilon$
transverse momentum, rather than as a function of event topology.

At low $\Upsilon$ transverse momentum, ATLAS finds that $\Upsilon(1S)$
events are accompanied by a larger charged-particle multiplicity than
$\Upsilon(2S)$ and $\Upsilon(3S)$ events, with absolute differences of
$3.6\pm0.4$ particles ($12\pm1\%$) and $4.9\pm1.1$ particles
($17\pm4\%$) respectively \cite{ATLASCONF2022023}. These differences
decrease with increasing $\Upsilon$ $p_T$, but do not vanish up to the
highest momenta studied ($p_T^{\mu\mu}\gtrsim 60$~GeV). This is
consistent with the CMS observation that $p_T$ selection weakens the
multiplicity dependence, while no published $p_T$ slice reproduces the
near-flat behaviour observed in the jet-like class ($S_T<0.55$).

This distinction between $p_T$ selection and topology selection is
essential for interpreting the two results together. Within CMS's own
data, the central slopes of the published $p_T$
slices remain negative, although the highest-$p_T$ interval is more
compatible with a flat behaviour, especially for $R_{21}$. None of the
published $p_T$ slices reproduces the near-flat behaviour observed in
the jet-like sphericity class. ATLAS, having measured the $p_T$
dependence only, is in effect independently confirming the ``$p_T$
alone does not flatten it'' side of the CMS picture. The two
measurements are therefore not in obvious tension: they probe
closely related associated-multiplicity observables under different
selections, one conditioned on $\Upsilon$ $p_T$ and the other on
event topology.

Three independent analyses address feed-down directly.
In CMS~2020 \cite{CMS2020}, the cone-isolation analysis explicitly
notes that feed-down from higher $\Upsilon$ states
($\Upsilon(2S)\to\Upsilon(1S)$, $\Upsilon(3S)\to\Upsilon(1S)$)
does not drive the cone dependence: no excess suppression is observed
in the dense-cone category ($N^{\Delta R}_{\rm track}>2$) relative to
the empty-cone category, independently of the feed-down composition.
CMS~2014 \cite{CMS2014pPb} discusses $\chi_b(nP)\to\Upsilon(nS)\gamma$
feed-down at length and concludes that conversion electrons from the
radiated photon have $p_T \lesssim 0.4$~GeV, below the reconstruction
threshold, so this dominant radiative channel cannot add charged
tracks.  ATLAS \cite{ATLASCONF2022023} ran \textsc{Pythia8} with and
without the implemented $\Upsilon(nS)\to\Upsilon(mS)$ feed-down chains,
and varied the colour-reconnection scheme: the observed
associated-multiplicity hierarchy between the three states is not
reproduced in any configuration.  Taken together, these three checks
indicate that no published feed-down implementation accounts for the
full pattern of observations.  A dedicated feed-down-aware calculation
would be the definitive test; none is currently available.

\section{Additional constraints from pPb and light-ion collisions}
\label{sec:ppb_constraints}

The $pp$ constraints of the preceding sections are complemented,
to our knowledge for the first time in a combined differential analysis,
by measurements in proton-lead collisions, which probe the same
suppression mechanism in a denser environment.

\paragraph{CMS pPb: evolution from 5.02 to 8.16 TeV.}
The event-activity dependence of $\Upsilon(nS)/\Upsilon(1S)$ in
pPb collisions was first measured by CMS at
$\sqrt{s_{_{\rm NN}}}=5.02$~TeV \cite{CMS2014pPb}, using both
charged-particle multiplicity and the forward transverse energy
$E_T(|\eta|>4)$ as activity estimators, with $\Upsilon$ states
reconstructed at $|y_{\rm CM}|<1.93$.  Both ratios decrease with
increasing $E_T$, establishing already in 2014 that the suppression
correlates with event activity measured several units of pseudorapidity
away from the quarkonium.

The CMS Physics Analysis Summary HIN-25-005 (preliminary)
\cite{CMSPAS2025} extends this programme to
$\sqrt{s_{{\rm NN}}}=8.16$~TeV with substantially larger statistics.
The two activity estimators used in 2014 --- charged-particle multiplicity
and forward $E_T(|\eta|>4)$ --- are both retained, now with a finer
binning in $E_T$ and with the addition of a cone-isolation study.
The preliminary 8.16~TeV result extends the 2014 measurement
\cite{CMS2014pPb} with higher statistics, finer $E_T$ binning, and
the addition of cone isolation; the long-range forward-$E_T$
correlation was already established in the published dataset.

Notably, the $\Upsilon(nS)/\Upsilon(1S)$ ratios as a function of
corrected $\Ntrack$ in pPb at 8.16~TeV are largely consistent with
the corresponding ratios measured in $pp$ at $\sqrt{s}=7$~TeV
\cite{CMS2020}, suggesting that charged-particle multiplicity is an
effective ordering variable across collision systems.
As a function of forward $E_T$,
$\Upsilon(2S)/\Upsilon(1S)$ falls from $\approx0.23$ at low $E_T$ to
$\approx0.15$ at $E_T\sim60$~GeV, while $\Upsilon(3S)/\Upsilon(1S)$
falls from $\approx0.13$ to $\approx0.05$ --- a substantially
larger relative decrease for the more weakly bound state,
confirming the sequential binding-energy hierarchy.

The two activity estimators --- midrapidity $\Ntrack$ and forward
$E_T(|\eta|>4)$ --- give consistent suppression trends despite probing
different pseudorapidity regions.  The pseudorapidity gap
$\Delta\eta \gtrsim 2$ between the $\Upsilon$ and the HF calorimeter
suppresses autocorrelation, and CMS concludes that the suppression
is not an artifact of the choice of estimator but a genuine physical
effect correlated with global event activity \cite{CMSPAS2025}.

Notably, a hint of a decreasing trend of $\Upsilon(3S)/\Upsilon(1S)$
with forward $E_T$ in $pp$ collisions at $\sqrt{s}=2.76$~TeV was
already visible in the CMS 2014 data \cite{CMS2014pPb}, with limited
statistics and few data points (reproduced in Fig.~3 of
Ref.~\cite{CMSPAS2025} for comparison).
This establishes that the forward-$E_T$ long-range correlation
was present in pp already in the 2014 dataset, independently of
the new pPb measurements at 8.16~TeV.

The cone-isolation study in HIN-25-005 compares events with no tracks
inside the cone ($N^{\Delta R}_{\rm track}=0$, ``isolated'') with events
having at least one track ($N^{\Delta R}_{\rm track}\geq1$).
The two cone categories give suppression ratios that are statistically
compatible within uncertainties for both $\Upsilon(2S)/\Upsilon(1S)$
and $\Upsilon(3S)/\Upsilon(1S)$.  The robust conclusion is the
\emph{absence} of the enhancement expected for local-density
dissociation \cite{CMSPAS2025}.

An explicit model comparison in HIN-25-005 comes to a nuanced
conclusion.  For the single ratios, a hot-medium framework based on a
time-dependent Schr\"odinger equation with a complex potential and a
hydrodynamic background (Chen et al.\ \cite{ChenQGP}) is found
consistent with the data across the full multiplicity range, with the
measured points lying within the predicted bands \cite{CMSPAS2025}.
By contrast, the Ferreiro--Lansberg comover-interaction model
\cite{FerreiroLansberg2018} fails to describe the single-ratio
multiplicity dependence and \emph{systematically overestimates} the
data across the full range, indicating that this published comover
implementation does not capture the absolute suppression of the
excited states in the preliminary pPb dataset \cite{CMSPAS2025}.
For the normalised double ratios, where the multiplicity normalisation
cancels much of the effect, the comover model reproduces the data at
higher multiplicity but lies below them at low multiplicity; CMS
attributes this change of trend largely to the cancellation inherent
in the normalised observable \cite{CMSPAS2025}.

No SHINCHON comparison is used in HIN-25-005 for the pPb
event-activity-dependent ratios.  The SHINCHON comparison belongs
instead to the light-ion analysis HIN-25-015, where CMS compares OO
and PbPb bottomonium double ratios with hydrodynamic-dissociation
predictions \cite{CMSPAS2025OO}.  It is therefore discussed separately
in the light-ion paragraph and in Section~\ref{sec:models}, not as
part of the pPb model comparison.

With the caveat that this is a preliminary dataset, one further point
is relevant for the comover model specifically, concerning an
observable it has not addressed at all.

\textit{Versus forward $E_T$:} to our knowledge, no prediction from
any of the frameworks discussed here --- CIM, partonic CIM, pNRQCD/OQS,
QGP droplet, or SHINCHON --- for the $\Upsilon(nS)$ suppression as a
function of $E_T(|\eta|>4)$ has been published.
The observed correlation between suppression and forward $E_T$
--- present in pPb at 5.02~TeV \cite{CMS2014pPb} and extended to
pPb at 8.16~TeV \cite{CMSPAS2025} --- therefore represents an
observable that the CIM framework has not yet addressed.
Whether an extended CIM implementation could reproduce this
long-range correlation is an open question.

\paragraph{LHCb pPb and Pbp: forward-backward asymmetry.}
\label{sec:decorrelation}
LHCb has measured the prompt $\psi(2S)/J/\psi$ ratio in both pPb
(proton-going, $1.5<y^*<4.0$) and Pbp (lead-going,
$-5.0<y^*<-2.5$) configurations at
$\sqrt{s_{_{\rm NN}}}=8.16$~TeV \cite{LHCbpPb2025}.
In the proton-going direction the prompt ratio decreases significantly
with multiplicity.  In the Pb-going direction, where a simple
monotonic density-based picture might have suggested a stronger
multiplicity dependence, the prompt ratio is instead suppressed but
approximately flat within uncertainties.

This forward-backward asymmetry constitutes a qualitative stress test
for density-driven pictures: the denser Pb-going configuration does
not show the steeper multiplicity dependence seen in the proton-going
direction.  LHCb explicitly concludes that the
suppression patterns in Pbp collisions cannot be fully explained by
comovers alone, and that the larger suppression of $\psi(2S)$ suggests
an additional mechanism, possibly linked to QGP-like effects
\cite{LHCbpPb2025}.  The pPb/Pbp comparison should therefore be read
as a qualitative stress test rather than a direct falsification of any
specific model.

The pattern in the Pb-going direction is consistent with suppression
having reached a saturation level, with the ratio varying little with
multiplicity.  Any viable model must reproduce both the monotonically
decreasing trend in pPb and the flat, strongly suppressed behaviour in
Pbp.

\paragraph{Light-ion collisions: OO and NeNe.}
A further extension of the system-size dependence is provided by the
recent CMS measurement of sequential $\Upsilon(nS)$ suppression in
oxygen-oxygen and neon-neon collisions at $\sqrt{s_{_{\rm NN}}}=5.36$~TeV
\cite{CMSPAS2025OO} (preliminary).  These symmetric light-ion systems
populate an intermediate system-size region
between $pp$/$p$Pb and PbPb, and CMS observes the same sequential
hierarchy: all excited-to-ground-state double ratios are below unity,
with the more weakly bound states more strongly suppressed
(e.g.\ $\Upsilon(2S)/\Upsilon(1S) = 0.664$ and
$\Upsilon(3S)/\Upsilon(1S) = 0.392$ in OO).  These double ratios are
normalised to the inclusive $pp$ reference at
the same centre-of-mass energy, not to a $pp$ sample selected at the
same event activity.  Since the inclusive $pp$ reference is dominated
by low-activity events, a double ratio below unity in OO or NeNe
primarily shows that the typical light-ion event samples a higher
activity environment than inclusive $pp$.  It should therefore not be
interpreted, by itself, as evidence for a qualitatively new
nucleus-specific suppression mechanism.

The more relevant comparison for the present constraint analysis is
provided by the single ratios plotted as a function of
$N_{\rm track}^{\rm corr}$.  In that representation, the $pp$, $p$Pb
and light-ion points follow a smooth common activity dependence, with
the low-activity OO and NeNe points lying close to the $pp$ values at
comparable multiplicity.  This is consistent with global event activity
being an important ordering variable, though the data do not by
themselves discriminate between global and local mechanisms.

The significance of this cross-system observation lies in the system
chosen: OO is a genuinely nuclear collision, in which one might
\emph{a priori} expect the onset of additional nuclear or deconfined
medium effects, yet at comparable event activity its single ratios
remain close to those measured in $pp$.  This supports the use of
event activity as a common ordering variable across systems, without
establishing by itself whether the underlying mechanism is local,
global, hadronic, pre-hadronic, or partonic.

CMS overlays SHINCHON predictions for the light-ion bottomonium
\emph{single} ratios as a function of $\Ntrack$ \cite{CMSPAS2025OO},
where the framework reproduces the qualitative trend.  In the double
ratio $\Upsilon(3S)/\Upsilon(2S)$ as a function of $p_T$, however,
the SHINCHON predictions do not give a satisfactory description of the
data.  In this framework, an event-by-event hydrodynamic description
of the medium is combined with bottomonium dissociation through
gluo-dissociation and inelastic parton scattering, including feed-down
contributions to the inclusive yields.

Taken together, the pPb results extend the constraint map to a denser
environment and add a direct non-locality argument: the correlation
between suppression and forward $E_T$ across a rapidity gap
$\Delta\eta \gtrsim 2$ (Constraint~5 of Section~\ref{sec:combined})
cannot be generated by local comover density at the quarkonium rapidity.
The light-ion results show that the single-ratio trend versus
corrected multiplicity is continuous across $pp$, $p$Pb, OO and NeNe,
supporting global event activity as an ordering variable; they do not
by themselves provide an independent argument against local-density
mechanisms.

\section{From early-time constraints to a dense coloured-medium interpretation}
\label{sec:partonic}

The multi-differential constraints establish \emph{what the suppression
mechanism is not}: not controlled by local charged-particle density
around the quarkonium, not azimuthally directed, not a function of
multiplicity alone, not purely rapidity-local, and not independent of
the finite spatial scale probed by transverse momentum.  In this section
we discuss the properties that the suppression mechanism appears to
require.  The argument proceeds in two steps.  The first step --- that
the relevant environment is pre-hadronic --- is supported by
formation-time arguments.  The second step --- that the early coloured
medium is specifically \emph{partonic} --- requires the density
argument and is supported only circumstantially by independent
observables.

\subsection{Step 1: Early-time coloured medium constraint}
\label{sec:prehadronic}

The suppression must act during the early separation phase of the
$b\bar{b}$ pair, before it has evolved into a physical bound state.
The $b\bar{b}$ pair is produced at the hard scattering and begins
to separate; it requires a formation time $\tau_{\rm form} \sim 1/(2m_b v^2)$
to evolve into a physical bound state, where $v$ is the relative
velocity of the quarks.
For the \ups{1} this gives $\tau_{\rm form} \sim 0.3$--$0.5$~fm;
for the more loosely bound \ups{3}, $\tau_{\rm form}$ is shorter
because $v$ is larger, of order 0.1--0.3~fm \cite{Brambilla2011}.
Light hadrons have formation times $\tau_{\rm form} \sim 1/m_\rho
\sim 1$~fm, substantially longer.
The CIM partonic implementation uses $\tau_0 \sim 0.1$~fm
\cite{FerreiroLansberg2018}, consistent with this pre-hadronic window.

At $\tau \lesssim 0.5$~fm, neither the quarkonium bound state nor
the surrounding light hadrons have yet formed.
The surrounding matter is therefore \emph{pre-hadronic} by kinematic
necessity --- it has not yet undergone string fragmentation into
colour-neutral hadrons.
This is a kinematic statement, independent of any dynamical assumption.

\textbf{Important caveat.}  Pre-hadronic does not mean partonic.
A pre-hadronic medium could consist of colour strings, flux tubes,
or glasma fields --- none of which are deconfined quark-gluon matter.
String-based models (ropes, colour reconnection) \cite{Sjostrand2018} also
operate in the pre-hadronic window and are therefore not excluded by
the timing argument alone.  The Color Glass Condensate (CGC)
\cite{McLerranVenugopalan1994,Gelis2010}, an initial-state gluon-saturation
framework rather than a hadronic or string-based final-state mechanism,
describes the glasma stage on a comparable timescale and is likewise
not distinguished from the proposed medium by timing alone.  The additional step requires an energy-density consistency argument.

\subsection{Step 2: Energy-density context}
\label{sec:deconfinement}

The pre-hadronic window identified in Step~1 ($\tau \lesssim 0.5$~fm)
coincides with the epoch of highest energy density in the collision.
Order-of-magnitude Bjorken estimates for high-multiplicity $pp$
collisions at LHC energies \cite{CsanadCsorgoJiangYang2017},
\begin{equation}
  \varepsilon_{\rm Bj} \sim \frac{dE_T/d\eta}{\tau_0 \, A_\perp},
\end{equation}
yield values in the range $\varepsilon \sim 1$--10~GeV/fm$^3$ at
$\tau_0 \sim 0.5$~fm and $A_\perp \sim 1$~fm$^2$, depending on the
assumed initial overlap area.
The QCD crossover energy density from lattice QCD is
$\varepsilon_c \approx 0.5$~GeV/fm$^3$ \cite{HotQCDLattice}.
The estimated energy densities at high multiplicity are therefore
comparable to or above this crossover scale, making a partonic
interpretation plausible.

\section{Relation to the Campanini--Ferri equation-of-state analysis}
\label{sec:eosconnect}

Campanini and Ferri \cite{CampaniniFerri2011} identified a qualitative
change in collective soft-sector behaviour in inclusive $pp$ and
$p\bar{p}$ multiplicity distributions, using an experimental
equation-of-state (EOS) proxy based on the interplay between charged-particle
density and mean transverse momentum --- built entirely from
soft-sector observables without any quarkonium input.
The central physical ingredient --- $\langle p_T\rangle$ as a temperature
proxy and multiplicity as an entropy proxy --- was subsequently formalised
by Gardim, Giacalone, Luzum and Ollitrault \cite{GardimOllitrault2020}.
Gardim, Giacalone and Ollitrault \cite{GardimGiacaloneOllitrault2020}
showed that the $\langle p_T\rangle$--multiplicity slope is proportional
to $c_s^2$, the squared speed of sound.  CMS has extracted
$c_s^2 = 0.241 \pm 0.002 \pm 0.016$ from ultra-central PbPb collisions
\cite{CMSSpeedOfSound2024}, and extended the approach to pPb
\cite{CMSPAS2025HIN25001}.

The multiplicity window in which the EOS proxy signals a qualitative
change in soft-sector behaviour --- broadly $dN_{\rm ch}/d\eta \sim 6$--20
from ISR to LHC energies \cite{CampaniniFerri2011} --- overlaps with the
window where the $\Upsilon(nS)/\Upsilon(1S)$ suppression ratios begin
their significant decrease with multiplicity in the CMS data.
This multiplicity scale corresponds to the Bjorken energy-density
range discussed in Section~\ref{sec:deconfinement}, where the
estimated $\varepsilon_{\rm Bj}$ becomes comparable to the QCD
crossover density.
The coincidence that both independent analyses converge on the same
characteristic multiplicity window, approached from entirely
different observables, is intriguing.
The EOS analysis predates the quarkonium measurements by several years;
the overlap is an independent cross-check, not a circular argument,
and is not used as input to the constraint analysis.

\subsection{What the differential constraints add: spatial structure}
\label{sec:diffadd}

The arguments above are consistent with the formation of an early, dense pre-hadronic coloured stage, with partonic degrees of freedom being a physically natural description at the estimated densities.
The multi-differential constraints of
Sections~\ref{sec:cone}--\ref{sec:ppb_constraints}
characterise its \emph{spatial structure}, which goes beyond the
question of pre-hadronic vs.\ partonic:

\begin{itemize}
  \item \textbf{Cone} (in $pp$ and $p$Pb): medium not concentrated
        locally around the quarkonium --- spatially extended.
  \item \textbf{Azimuthal symmetry} (in $pp$): medium not directionally
        aligned with the quarkonium momentum.
  \item \textbf{Sphericity} (in $pp$): medium density is event-shape
        dependent --- isotropic events produce a denser, more uniform medium.
  \item \textbf{$p_T$ dependence} (in $pp$): the progressive weakening
        of the multiplicity dependence with increasing $\Upsilon$ $p_T$
        is qualitatively consistent with a suppression region of finite
        spatial extent $R$, since a high-$p_T$ quarkonium traverses
        it in a shorter proper time $\tau_{\rm cross}\sim R/(\beta\gamma)$.
  \item \textbf{$E_T$ gap} (in $p$Pb; hint in $R_{31}$ in $pp$):
        medium globally correlated over large rapidity intervals,
        not localised at the quarkonium rapidity.
  \item \textbf{Possible threshold-like behaviour in the Pb-going direction}:
        the suppression in the Pb-going configuration shows a pattern
        potentially consistent with threshold-like behaviour rather than
        a simple smooth density scaling, though quantitative modelling is
        required before stronger conclusions can be drawn.
\end{itemize}

\section{Falsifiable predictions}
\label{sec:predictions}

The differential constraints established here imply several falsifiable
predictions that future measurements can test to discriminate between
mechanisms satisfying the constraint set.

\paragraph{Survival anisotropy in non-central collisions.}
The azimuthal-sector equivalence established in
Section~\ref{sec:azimuth} shows that, in the inclusive $pp$ sample,
the suppression is not driven by charged-particle activity in a
specific azimuthal sector relative to the reconstructed $\Upsilon$
direction.
This does not exclude a path-length dependence in systems where an
initial geometric asymmetry exists.
In non-central $p$Pb or peripheral AA collisions, where the overlap
region is not azimuthally symmetric, a path-length-dependent
suppression mechanism would predict a binding-energy-ordered elliptic
flow hierarchy:
\begin{equation}
v_2(\Upsilon(3S)) > v_2(\Upsilon(2S)) > v_2(\Upsilon(1S)).
\end{equation}
A hierarchy of this kind was predicted within the pNRQCD/Lindblad
framework by Islam and Strickland \cite{IslamStrickland2021} for PbPb
collisions. A measurement of $v_2[\Upsilon(2S)+\Upsilon(3S)] -
v_2[\Upsilon(1S)]$ in non-central $p$Pb would test whether the
path-length dependence of the suppression is consistent with the
finite-size interpretation established here, without contradicting
the azimuthal isotropy observed in the inclusive $pp$ sample.

\paragraph{CIM cone and sphericity calculations.}
The CIM, including its early partonic/pre-hadronic implementation, has not yet
been applied to cone-isolated or
sphericity-selected samples.  An explicit calculation reproducing both
$R_{21}(N_{\rm track})$ and $R_{21}(N_{\rm track}|S_T)$ at fixed
multiplicity would constitute a strong test of whether the published
azimuthally averaged density profile can be modified to pass the
topology constraint.

\paragraph{Cone isolation in pPb.}
The CMS preliminary cone-isolation result in pPb \cite{CMSPAS2025}, if confirmed
by the final publication, should be measured with full statistics and compared
quantitatively with the $pp$ result at matched multiplicity.  At the level of
central values, HIN-25-005 shows no enhancement of suppression in the
non-isolated category, and for $R_{21}$ the isolated category is even more
suppressed.  The robust present conclusion is therefore the absence of the
positive local-density ordering expected from a simple local comover picture.
A final result confirming either cone-independence or inverse cone-ordering
would constitute a strong constraint on local-density dissociation mechanisms.

\paragraph{Topology-resolved charmonium at LHCb.}
The LHCb measurement of $\psi(2S)/J/\psi$ vs multiplicity in $pp$ at
13~TeV \cite{LHCbPsi2024} could in principle be extended to
sphericity-selected subsamples.  If the same isotropic/jetty splitting
observed for bottomonium is present for charmonium, it would validate
the universality of the topology constraint across quarkonium species.

\paragraph{Triple-differential CMS measurement: $R_{n1}(N_{\rm track},S_T,p_T)$.}
The present analysis combines the $p_T$-sliced and sphericity-sliced
CMS data separately, but the two selections have not been applied
simultaneously.  A measurement of $R_{21}$ and $R_{31}$ as a function
of $N_{\rm track}$ in bins of both $S_T$ \emph{and} $p_T$
simultaneously would directly answer whether the sphericity dependence
survives at fixed $p_T$, or whether it is fully accounted for by the
different $p_T$ compositions of jet-like and isotropic event classes.
This triple-differential measurement is currently absent from the
published literature and is identified as the most direct available
test to separate the topology effect from the kinematic $p_T$ bias
discussed in Section~\ref{sec:sphericity}.

\section{Interpretation and conclusions}
\label{sec:conclusions}

The analysis presented here combines the six complementary
differential constraints introduced in Section~\ref{sec:combined}.
Each observable constrains a different physical property of the
suppression mechanism, and the combined set is more restrictive than
the inclusive multiplicity dependence alone.

The positive conclusions are stated at three levels of inference.

\textbf{Level 1 --- direct constraints from data}:
\begin{itemize}
  \item A suppression mechanism controlled only by the local
        charged-particle density around the quarkonium is disfavoured
        (in $pp$ and $p$Pb): the multiplicity-dependent suppression
        trend is statistically indistinguishable across cone-isolation
        categories.
  \item A purely multiplicity-driven picture is disfavoured (in $pp$):
        transverse sphericity at fixed $\Ntrack$ separates jet-like
        and isotropic events.
  \item Directional or near-side explanations are disfavoured (in $pp$)
        by the compatibility of forward, transverse and backward
        azimuthal sectors.
  \item The multiplicity dependence of the suppression weakens
        progressively with increasing $\Upsilon$ $p_T$ (in $pp$;
        consistent with the ATLAS and LHCb observations at 13~TeV).
  \item A purely rapidity-local mechanism is disfavoured (in $p$Pb
        at 5.02 and 8.16~TeV; hint from $R_{31}$ in $pp$) by the
        correlation of suppression with forward $E_T$ across a
        large rapidity gap.
  \item A simple monotonic comover-density scaling is stressed by the
        contrasting pPb and Pbp multiplicity dependences observed by
        LHCb.
\end{itemize}

\textbf{Level 2 --- strong inference} (follows from Level 1 jointly):
any viable mechanism must be global, topology-dependent, and finite-size.
When combined with the quarkonium formation-time argument, the
suppression must act in an early pre-hadronic environment.  To our
knowledge, no published framework has been explicitly tested against
the full set of these differential constraints simultaneously.

\textbf{Level 3 --- interpretive context} (compatible with, not proven
by, the data): the constraints are compatible with a dense coloured
pre-hadronic medium, active at early times ($\tau \lesssim 0.5$~fm). This interpretation is supported by
independent context from other observables in the same multiplicity
window, each of which has alternative explanations and none of which
constitutes a proof of deconfinement in isolation:
\begin{itemize}[noitemsep,topsep=2pt]
  \item strangeness enhancement ($\Xi$, $\Omega$) at
        $dN_{\rm ch}/d\eta \gtrsim 10$--15 (ALICE
        \cite{ALICEStrangeness2017,ALICEStrangenessExtremes2025});
  \item long-range azimuthal ridge correlations at similar
        multiplicities (CMS \cite{CMSRidge2010}); a recent ALICE
        measurement extends the ridge signal down to
        $\langle N_{\rm ch}\rangle \gtrsim 9$ (charged particles
        in $|\eta|<1$, $p_T>0.2$~GeV/$c$) in $pp$ collisions at
        $\sqrt{s}=13$~TeV, and demonstrates that the ridge yield
        in pp exceeds the limits set in $e^+e^-$ annihilations
        by 3.8--5$\sigma$ in the range
        $8 \lesssim \langle N_{\rm ch}\rangle \lesssim 24$,
        indicating that the correlations originate from processes
        specific to hadronic collisions \cite{ALICERidgeLowMult2024};
  \item partonic flow signatures: ALICE observes a distinctive
        baryon--meson $v_2$ grouping and splitting (${\sim}5\sigma$)
        at intermediate $p_T$ in high-multiplicity $pp$ and $p$Pb
        collisions, selected to have the same average charged
        multiplicity, $\langle N_{\rm ch}\rangle \approx 35$
        (for $0.2<p_T<3.0$~GeV/$c$, $|\eta|<0.8$)
        \cite{ALICEPartonicFlow2026}, reproduced only by
        a model including hydrodynamic evolution followed by quark
        coalescence (Hydro-Coal-Frag); CGC, hadronic rescattering,
        rope hadronisation, and AMPT all fail to reproduce the
        baryon--meson splitting \cite{ALICEPartonicFlow2026}.
        The earlier CMS $v_2$ mass ordering \cite{CMSFlow2017}
        is consistent with the same picture.
  \item extreme strangeness, with $\Omega/\pi$ approaching values
        typical of heavy-ion collisions (ALICE \cite{ALICEStrangeness2017});
        a recent ALICE measurement extends this to the regime of multiple
        multi-strange hadrons per event, reaching strangeness-content
        imbalances $\Delta S=5$ with enhancements of two orders of
        magnitude from low to high multiplicity \cite{ALICEStrangenessExtremes2025}.
\end{itemize}

This interpretive context is coherent with two further considerations.
First, the multiplicity window in which these signatures appear is the
same one in which, as discussed in Sections~\ref{sec:deconfinement}
and~\ref{sec:eosconnect}, the estimated Bjorken energy density at
$\tau \lesssim 0.5$~fm becomes comparable to or above the QCD crossover
scale, and in which the soft-sector equation-of-state proxy of
Campanini and Ferri signals a qualitative change.  Second, the absence
of a clear jet-quenching signal in high-multiplicity $pp$ is not in
contradiction with this picture: radiative energy loss scales as
$\Delta E_{\rm rad}\propto\hat{q}L^2$ (collisional loss closer to
linear in $L$), so the expected jet-quenching signal in a small system with
$L\sim1$--2~fm is strongly suppressed compared to a central heavy-ion
collision with $L\sim5$--6~fm, while the few-hundred-MeV binding of $\Upsilon(3S)$
can still be modified at a much smaller scale than that needed for a
robust high-$p_T$ quenching signal.  In this light, the present
sequential-suppression pattern is precisely the kind of observable that
Sj\"ostrand identifies as a key open question for high-multiplicity
$pp$ collisions, alongside the question of which features constitute
ironclad signatures of QGP formation \cite{Sjostrand2018}; the
constraint set established here is intended as experimental input to
that question.

The present work does not identify a unique surviving model, nor does
it claim that any individual observable constitutes evidence for
deconfinement.  Its contribution is to define a common experimental
benchmark --- the simultaneous set of differential constraints
introduced above --- that any successful microscopic description must
reproduce at once: an early, global, geometry-sensitive, and
finite-size mechanism.  A dense, coloured, early-time stage is a
natural candidate, but alternative frameworks remain viable provided
they satisfy the full constraint set; the data make a conventional
local-density, multiplicity-only, or purely late-hadronic interpretation
difficult to maintain.

Among these constraints, the scissors constraint --- the local-density
picture disfavoured by the cone test, the pure-multiplicity picture
disfavoured by the sphericity test --- is the conceptually most
original element and the one most robustly supported by the data. It
rests on the published CMS 7~TeV \cite{CMS2020} and LHCb
\cite{LHCbPsi2024} measurements alone; the preliminary HIN-25-005
\cite{CMSPAS2025} and light-ion \cite{CMSPAS2025OO} data are used only
as supporting cross-system evidence. Dedicated calculations within the
CIM and other viable frameworks, applied to the cone-selected,
azimuth-resolved, sphericity-resolved, and long-range correlation
observables discussed here, would establish which mechanisms can
reproduce the full constraint set.

The robustness of the overall constraint map against the
$p_T$-composition caveat on the sphericity observable, and against
possible feed-down contributions, deserves an explicit note.
Even if part of the observed topology dependence were ultimately found
to originate from a different $p_T$ composition of jet-like and
isotropic events, or if feed-down were found to contribute more
substantially than current estimates suggest, the cone-isolation
constraint would remain unchanged, as would the forward-$E_T$
long-range correlation and the pPb/Pbp forward-backward asymmetry.
The overall constraint map would therefore survive, although the
interpretation of the sphericity observable might require revision.
The scissors constraint in its strongest form would be weakened, but
the remaining constraints would still require an early, globally
correlated, finite-size suppression mechanism.

\section*{Data availability}

All numerical information used in the analysis is taken from publicly available CMS, LHCb and ALICE publications and their associated public data releases where available. In addition, the preliminary CMS Physics Analysis Summaries HIN-25-005 \cite{CMSPAS2025} and HIN-25-015 \cite{CMSPAS2025OO} are used as supporting evidence for the $p$Pb and light-ion constraints discussed in Section~\ref{sec:ppb_constraints}; results based on these preliminary datasets are explicitly identified as such throughout the text and are not required for the core conclusions of the paper. The figures included here are based on the public-data plots used in the submitted analysis. No new experimental measurement is reported.

\appendix
\section{Jet-like class: comparison with $p_T$-sliced CMS measurements}
\label{app:slopes}

This appendix provides a quantitative comparison between the jet-like
sphericity class ($S_T<0.55$) and the published $p_T$-sliced CMS
projections, using weighted linear fits
$R_{n1}(\Ntrack) = a_{n1} + b_{n1}\,\Ntrack$ to the CMS HEPData
central values and total uncertainties \cite{CMS2020}.
Table~\ref{tab:ptmatch} gives the fit parameters for the seven $p_T$
slices and the two extreme sphericity classes.

\begin{table}[h]
\centering
\begin{tabular}{lcccc}
\hline
Selection & $a_{21}$ & $b_{21}\;[10^{-3}\,\Ntrack^{-1}]$ &
  $a_{31}$ & $b_{31}\;[10^{-3}\,\Ntrack^{-1}]$ \\
\hline
$0$--$5$~GeV/$c$    & 0.2498 & $-0.99\pm0.25$ & 0.1211 & $-0.54\pm0.15$ \\
$5$--$7$~GeV/$c$    & 0.2949 & $-1.62\pm0.20$ & 0.1521 & $-1.16\pm0.13$ \\
$7$--$9$~GeV/$c$    & 0.3109 & $-1.24\pm0.12$ & 0.1778 & $-1.18\pm0.11$ \\
$9$--$11$~GeV/$c$   & 0.3395 & $-1.41\pm0.15$ & 0.2143 & $-1.49\pm0.13$ \\
$11$--$15$~GeV/$c$  & 0.3823 & $-1.31\pm0.15$ & 0.2416 & $-1.60\pm0.14$ \\
$15$--$20$~GeV/$c$  & 0.4276 & $-0.97\pm0.18$ & 0.2888 & $-1.52\pm0.16$ \\
$20$--$50$~GeV/$c$  & 0.4696 & $-0.34\pm0.25$ & 0.3496 & $-1.15\pm0.21$ \\
\hline
jet-like ($S_T<0.55$)  & 0.3493 & $-0.06\pm0.24$ & 0.2184 & $-0.48\pm0.20$ \\
isotropic ($S_T>0.85$) & 0.3224 & $-0.76\pm0.13$ & 0.2047 & $-0.99\pm0.10$ \\
\hline
\end{tabular}
\caption{Intercept $a_{n1}$ and slope $b_{n1}\pm\sigma_b$ from weighted
linear fits to the CMS HEPData points \cite{CMS2020}. Slope
uncertainties are from the fit covariance matrix.}
\label{tab:ptmatch}
\end{table}

The jet-like class has intercepts $a_{21}=0.349$ and $a_{31}=0.218$,
matching the $9$--$11$~GeV/$c$ normalisation range for both ratios,
while its slopes $b_{21}=-0.06\times10^{-3}$ and
$b_{31}=-0.48\times10^{-3}$ are near-zero. No published $p_T$ slice
simultaneously reproduces both intercept and slope for either ratio:
the slice best matching the intercept ($9$--$11$~GeV/$c$) has slopes
discrepant at $4.9\sigma$ ($R_{21}$) and $3.5\sigma$ ($R_{31}$);
the slice best matching the slope ($20$--$50$~GeV/$c$) has intercepts
discrepant at $5.0\sigma$ ($R_{21}$) and $4.8\sigma$ ($R_{31}$)
(joint test $T^2=t_a^2+t_b^2$, diagonal approximation,
$p<10^{-5}$ for every slice and both ratios).
Within the published one-dimensional projections, the $p_T$ composition
does not provide a natural explanation for the near-vanishing slopes of
the jet-like class; a definitive separation requires the
triple-differential measurement proposed in
Section~\ref{sec:predictions}.

\section*{Acknowledgement of AI assistance}

AI-based tools were used for editorial assistance, \LaTeX{} restructuring
and consistency checks during manuscript preparation.
The scientific interpretation, responsibility for the analysis and
final text remain with the author.

\end{document}